\documentclass[aps,prb,reprint,superscriptaddress]{revtex4-2}
\usepackage{hyperref}
\usepackage{graphicx}
\usepackage{color}
\usepackage{amsfonts}
\usepackage{amsmath}
\usepackage{physics}

\usepackage[utf8]{inputenc}
\usepackage[T1]{fontenc}
\usepackage{mathtools}
\usepackage{bbm}
\usepackage[dvipsnames]{xcolor}
\hypersetup{colorlinks=true, urlcolor=blue, citecolor = blue}
\usepackage[english]{babel}

\begin{document}

\title{Aharonov-Bohm interference probing \\of chiral Andreev edge states via scanning gate microscopy}

\author{S. Maji}
\email{maji@agh.edu.pl}
\affiliation{AGH University of Krakow, Academic Centre for Materials and Nanotechnology, al. A. Mickiewicza 30, 30-059 Krakow, Poland}

\author{M. P. Nowak}
\email{mpnowak@agh.edu.pl}
\affiliation{AGH University of Krakow, Academic Centre for Materials and Nanotechnology, al. A. Mickiewicza 30, 30-059 Krakow, Poland}

\date{\today}

\begin{abstract}
We theoretically investigate the spatial response of chiral Andreev edge states at a quantum Hall-superconductor interface probed by scanning gate microscopy. Using the Bogoliubov-de Gennes formalism, we demonstrate that the localized potential of the scanning gate tip decouples the chiral Andreev edge states from the superconductor into independent electron and hole paths. In the $\nu=2$ regime, this local separation forces the quasiparticles to accumulate a relative magnetic phase in the normal region, resulting in Aharonov-Bohm-type conductance oscillations that are captured by a geometric analytical model. At higher filling factors (e.g., $\nu=4$), the tip progressively interacts with spatially distinct edge channels, dynamically altering the Andreev-induced mode mixing and modifying the transport resonances. Furthermore, we show that these highly localized interferometric signatures survive in the presence of disorder, even when conventional non-local transport oscillations are masked.
\end{abstract}

\maketitle
\section{Introduction}
The coupling between chiral edge states formed in a quantum Hall (QH) device with induced superconductivity (SC) sets up an interesting platform for studying the interplay between magnetic field, electron-hole conversion, and topology \cite{Anjana,Lee2017,Ronen, maji2026, Hoppe}. In a normal-superconductor (NS) interface, an incident electron with energy less than the superconducting gap experiences Andreev reflection and is reflected back as a hole, while a Cooper pair is transferred to the superconducting region \cite{osti_4071988}. When a QH device is coupled to a superconductor, the repeated Andreev conversion between electron-like and hole-like components gives rise to chiral Andreev edge states (CAESs) that propagate along the QH-SC interface. Quantum Hall-superconductor devices not only create a platform to explore the proximity effect in the QH regime, but the hybrid CAESs formed at the interface also attract considerable interest for their possible applications in topological superconductivity and chiral Majorana modes \cite{Chamon, Gamayun,Gaurav, Rakesh, Baba}.\\

The magnetic field required to establish the quantum Hall effect often destroys superconductivity; therefore, building a QH-SC hybrid structure is particularly challenging. Nevertheless, superconducting correlations were successfully incorporated in graphene based devices in the QH regime, where the superconducting proximity effect, Andreev reflection, and Josephson transport were observed \cite{Calado2015, BenShalom2016, Zhao2020, Borzenets}. Later on, a marriage of two-dimensional electron gas (2DEG) heterostructures with high critical field superconductors allowed the study of semiconducting-superconducting hybrids in the QH regime \cite{TAKAYANAGI1998462,Wan2015, Eroms}. The advances in the creation of highly transparent QH-SC interfaces motivated the detailed study of the transport phenomena of CAESs \cite{Lee2017, TAKAYANAGI1998462, Eroms, Zhao2020, Calado2015, BenShalom2016, Cuozzo, Hatefipour2022, Wan2015}. It was found that the interference of CAESs propagating along the QH interface results in conductance oscillations \cite{Zhao2020, Gamayun, maji2026} when parameters such as chemical potential, magnetic field, or propagation length are varied. Those oscillations are considered a significant signature of CAES. Although the transport measurements reveal significant signatures of CAESs, they do not allow for uncovering the presence of CAESs directly, and very often the conductance oscillations themselves are masked by disorder-induced noise \cite{Antonio}. 

Here we theoretically study scanning gate microscopy (SGM) as a tool to detect CAESs. In SGM experiments, a charged atomic force microscope tip acts as an electrostatic local gate. It moves throughout the system, creating a local perturbation that alters the charge carrier trajectories while the conductance change is registered. SGM has been used successfully to image and manipulate electron trajectories \cite{PhysRevB.77.125310, Martins, Pala_2009, Aidala2007, Topinka2000, Topinka2001, Topinka2009, Iagallo, LeRoy, LeRoy2002, Nowak2014, Prokop, Aoki, Nowak2014}, particularly in the quantum Hall regime \cite{PhysRevB.96.195423}. SGM studies have also been extended to NS hybrid structures. Theoretical studies have proposed the possible application of SGM in monitoring the modified trajectories of Andreev reflected holes in a NS junction \cite{S.Maji}, the supercurrent distribution in the Josephson junction \cite{Kaperek}, and providing signatures of Majorana bound states in a superconducting nanowire \cite{Maji}. Importantly, bent electron-hole orbits \cite{Bhandari2020} and, more recently, local modulation of the supercurrent have been experimentally observed by SGM \cite{Lombardi2025, villani2026scanninggatemicroscopymodulation}.

In this work, we theoretically study the SGM technique to uncover the spatial response of CAESs formed at the interface of a QH device and a superconductor. In normal QH systems, the electrons propagate unidirectionally at the edges, and the SGM tip cannot induce backscattering unless it connects the two counter-propagating edge states at opposite edges of the system. We show that in a QH-SC device, the SGM tip is capable of decoupling the CAESs from the superconductor. Therefore, when the SGM tip approaches the interface, the electron- and hole-like components of CAESs are separated locally and forced to move around the potential induced by the tip. Due to this detour, electrons and holes gain a relative phase, and Aharonov-Bohm-type interference is observed as an oscillating signature of CAESs in the system's conductance. We explain these oscillations using an analytical model that aligns with the numerical results and takes into account the modification of the induced electron-hole phase difference by the SGM tip. In addition, we investigate the system with higher filling factors and the possibility of using SGM to detect CAESs in non-ideal, disordered systems.

The paper is organized as follows. In Section II, we introduce our theoretical model. Section III.A presents the results for the lowest filling factor, along with an analytical explanation of the observed conductance oscillations. In Section III.B, we consider the case of a higher filling factor, and Section III.C describes the SGM mapping in disordered systems. The paper is concluded in Section IV.

\section{Theory}
\begin{figure}
    \centering
    \includegraphics[width=0.98\linewidth]{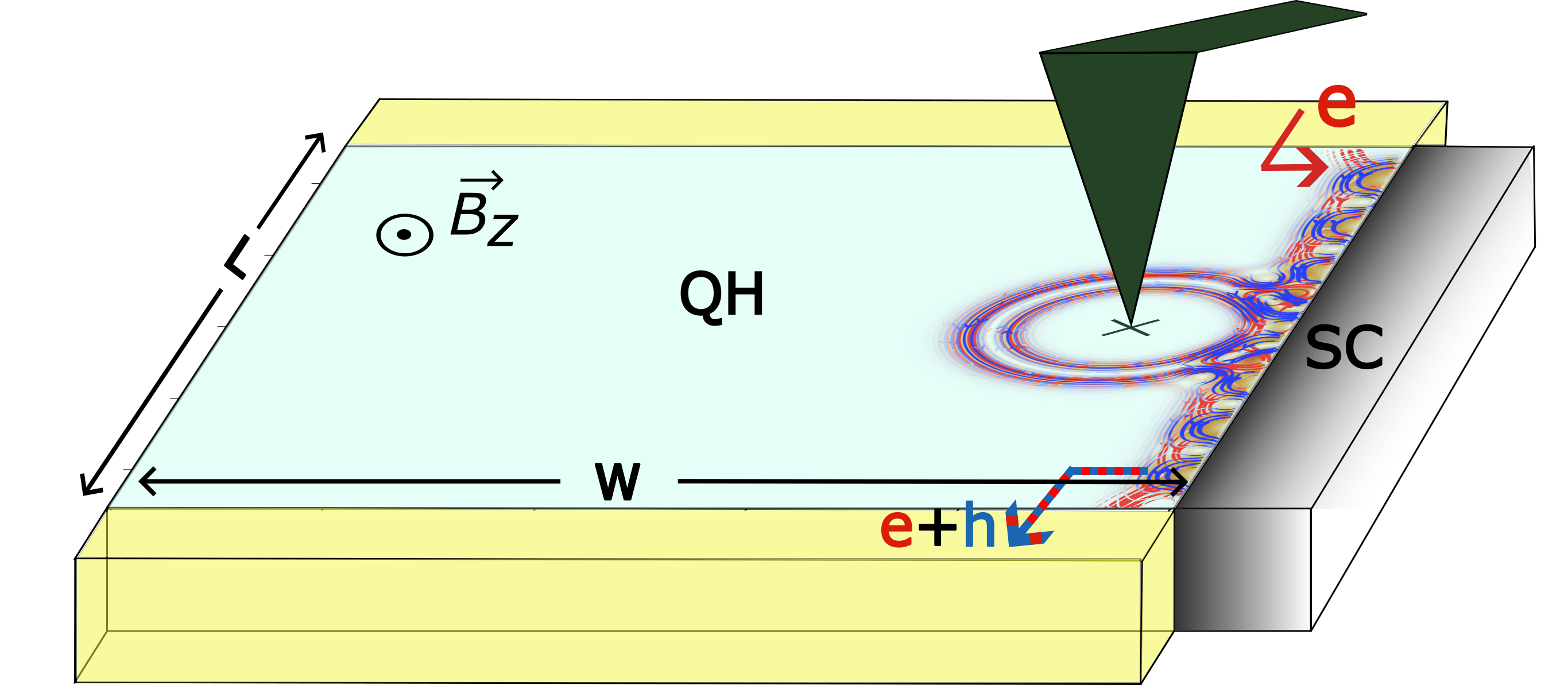}
    \caption{Schematic diagram of the QH-SC system. The light green part denotes the QH part, the gray region denotes the superconducting part and the yellow parts denote the semi-infinite normal leads. The electrons propagating on the QH edge mode forms CAES at the NS interface. We plot resulting probability current at the edge, where colors correspond to its electron (red) and hole (blue) character. The SGM tip, introduced in the system and denoted by the dark green color, locally perturbs the CAES at the QH-SC edge, detouring its path from the interface.} 
    \label{fig:System}
\end{figure}

The schematic of the considered device is shown in Fig.~\ref{fig:System}. It consists of a normal region where the perpendicular field induces a QH phase (light green) and a superconducting region (gray). The system is described by the Bogoliubov-de Gennes Hamiltonian
\begin{equation}
\begin{split}
\label{eqn:Hamiltonian_Zeeman_SO}
H = &\left(\frac{\hbar^2 \mathbf{k}^2}{2m^*} +V_{\mathrm{SGM}}(x,y, x_{\mathrm{tip}}, y_{\mathrm{tip}}) - \mu\right)\tau_z + \Delta(r)\tau_x,  
\end{split}
\end{equation}
where $\mathbf{k} = -i\nabla_r$, $\tau_x$, and $\tau_z$ are the $x$ and $z$ Pauli matrices acting on the wave function written in the basis $\Psi = (\Psi_e, \Psi_h)^T$. $\mu$ is the chemical potential, and $\Delta(r)$ is the superconducting pairing potential, non-zero in the superconducting segment. The perpendicular magnetic field is introduced by the substitution $\hbar k \rightarrow \hbar k - q\textbf{A}(r)$, where $\textbf{A}$ is the magnetic vector potential in the Landau gauge $\textbf{A} = (0, B_zx, 0)$ in the normal part of the system. We assume negligible Zeeman interaction strength and consider the system to be spin degenerate. In the tight-binding description, we discretize the Hamiltonian on a square lattice with a lattice constant $a$, and the orbital effects are implemented by adding a gauge-dependent phase factor in the hopping term $t_{ij}\rightarrow t_{ij}\exp[\frac{\mp ie}{\hbar}\int_{r_j}^{r_i} \textbf{A} \cdot d\vec{l}]$, with the sign being opposite in the electron/hole sector. We set the dimensions of the system to $W = 1000$ nm and $L = 1000$ nm unless stated otherwise and set $\Delta = 2$~meV.

To model the SGM potential at the 2DEG level, we use a Lorentzian potential distribution after Ref. \cite{Szafran}
\begin{equation}
V_{\mathrm{SGM}}(x,y)=\frac{V_{\mathrm{tip}}}{1+\frac{(x-x_{\mathrm{tip}})^2+(y-y_{\mathrm{tip}})^2}{d_{\mathrm{sgm}}^2}},
\end{equation}
where we set $V_\mathrm{tip}=0.01$ eV and $d_{\mathrm{sgm}}=50$ nm. The position of the SGM tip is determined by the pair of coordinates ($x_{\mathrm{tip}},y_{\mathrm{tip}}$).

To analyze the transport properties of the system, we adopt the Landauer-B\"uttiker approach. We consider two normal, semi-infinite leads connected at the top and bottom of the system (see yellow regions in Fig.~\ref{fig:System}). The conductance of the system is obtained via the formula:
\begin{equation}
G_{ij}(E) = \frac{\partial I_i}{\partial V_j} = \frac{2e^2}{h}(\delta_{ij}N_i^e(E)-T^{ee}_{ij}(E)+T^{he}_{ij}(E)).
\label{conductanceformula}
\end{equation}
$I_i$ is the current entering the scattering region from the terminal $i$, and $V_j$ is the voltage applied to the $j$'th lead. $N_i$ is the number of electronic modes in the $i$'th lead; $T^{ee}_{ij}(E)$ is the electron-to-electron transmission coefficient for electrons injected from the lead $j$ and captured at the terminal $i$; and $T^{he}_{ij}(E)$ is the corresponding electron-to-hole transmission coefficient. Here, we focus on  zero-temperature non-local conductance by placing $i \ne j$ and considering injecting the electrons from the top lead such that they create edge modes on the right side of the system in a positive perpendicular magnetic field. The corresponding calculated conductance becomes $G = G_{\mathrm{bottom}, \mathrm{top}}(E)$.

The coefficients $T^{ee}_{ij}(E)$ and $T^{he}_{ij}(E)$ are obtained from the scattering matrix of the system calculated at the energy $E$ corresponding to the voltage bias on the $j$'th lead,
\begin{equation}
    T^{\alpha,\beta}_{kl}(E) = \mathrm{Tr}\left( [S^{\alpha,\beta}_{kl}(E)]^\dagger S^{\alpha,\beta}_{kl}(E)\right).
\end{equation}
$S^{\alpha,\beta}_{kl}(E)$ is the block of the scattering matrix corresponding to the particles of type $\beta$ injected from the $l$'th lead and scattered back as the particle type $\alpha$ into the $k$'th lead. We particularly focus on a situation with negligible voltage bias and set $E=0$. 

The scattering matrix of the system is obtained by discretizing the Hamiltonian Eq. (\ref{eqn:Hamiltonian_Zeeman_SO}) on a mesh with lattice spacing $a = 5$ nm and solving the transport problem using the Kwant package \cite{Christoph}. Conductance maps are calculated with the help of the Adaptive Python package \cite{nijholt_2023_10215599}. The code used to obtain the results presented in this paper is available in an online repository \cite{maji_2026_22896864}.

\section{Results}
\subsection{Lowest filling factor case}
\begin{figure}
    \centering
    \includegraphics[width=0.85\linewidth]{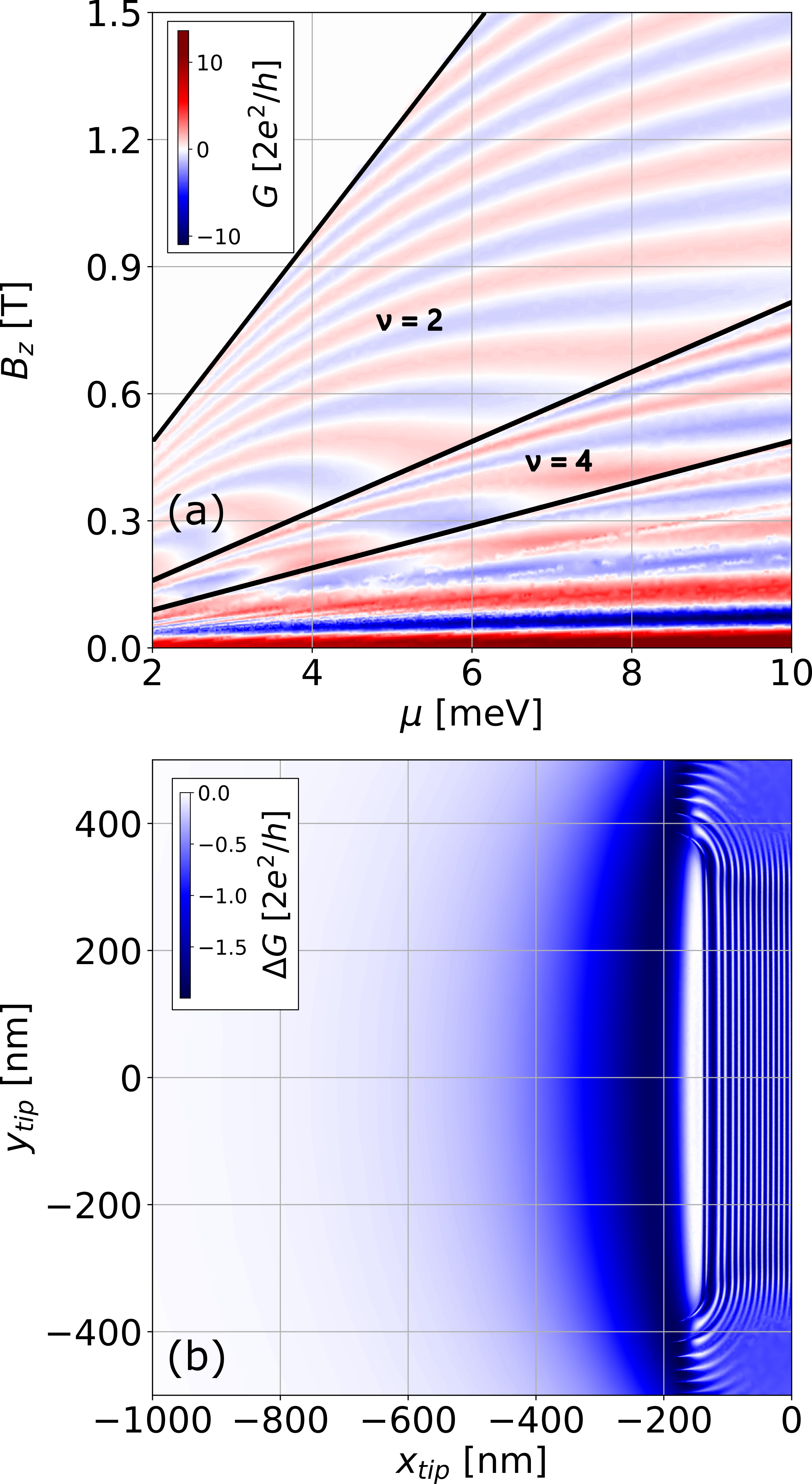}
    \caption{(a) Conductance map for QH-SC system without the SGM tip versus the chemical potential and the magnetic field. (b) Conductance difference varying the SGM tip position for $B = 0.912$ T, $\mu = 5$ meV.}
    \label{fig:Conductance_QH_SC_SGM_spinless}
\end{figure}

We start by inspecting the conductance of a pristine system, without the SGM tip. Figure \ref{fig:Conductance_QH_SC_SGM_spinless}(a) shows the non-local conductance with respect to the perpendicular magnetic field and the chemical potential. Black lines demarcate regions of the first few filling factors. We observe that as the chemical potential is increased and the value of the magnetic field decreases, the number of edge modes increases, amplifying the conductance. The map shows oscillating conductance patterns that are due to interference between CAESs, as is to be expected for clean QH-SC systems \cite{PhysRevB.93.161401, Zhao2020, maji2026}.

Let us first focus on the case of a single (electron, spin degenerate) edge mode, i.e. $\nu = 2$. Figure \ref{fig:Conductance_QH_SC_SGM_spinless}(b) shows the map of the conductance change introduced by the SGM tip, $\Delta G = G(x_\mathrm{tip}, y_{\mathrm{tip}}) - G_0$, where $G_0$ is the conductance obtained without the tip. As we selected the value of the pristine conductance close to the conductance quantum, the SGM tip can only decrease the system's conductance, represented by the overall blue color in the map. 

The map presents a unique feature of CAESs as probed by SGM: as the tip moves towards the edge on which the CAESs are located, the conductance starts varying until it begins to oscillate rapidly. This is in striking contrast to the case of normal QH edge states, whose topological protection from backscattering prevents any change in conductance until the disturbance is strong enough that it can connect the two counter-propagating edges of the sample \cite{PhysRevB.96.195423}.

\begin{figure}
    \centering
    \includegraphics[width=0.85\linewidth]{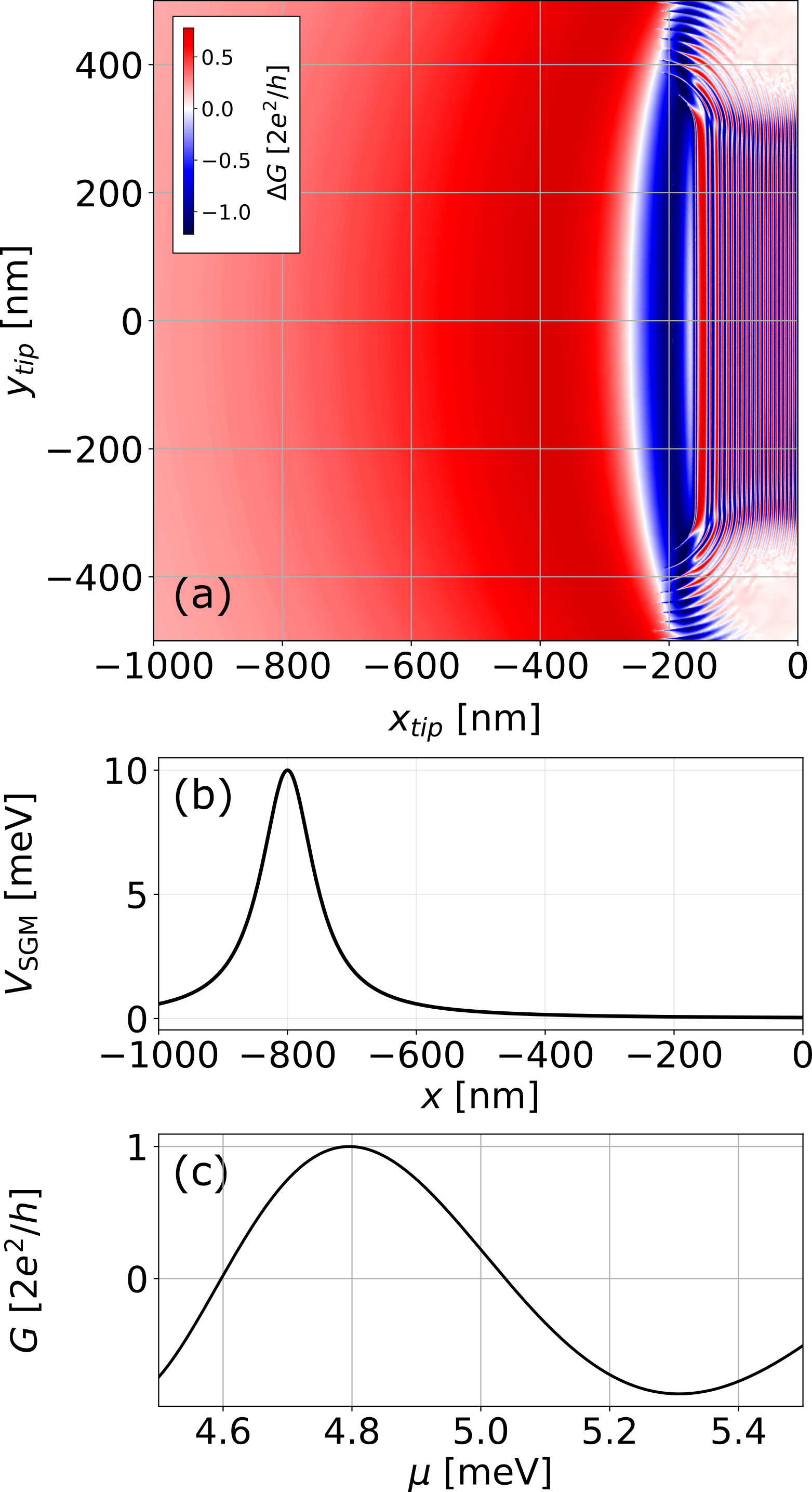}
    \caption{(a) Conductance difference varying SGM tip at $\mu = 5$ meV. (b) SGM potential distribution along the $x$-direction for the tip center located at $x = -800$ nm. (c) Conductance without the tip versus the chemical potential. The results are obtained for $B = 1$ T.}
    \label{fig:G_and_SGM_pot_dist}
\end{figure}

As the magnetic field is varied, the conductance of the pristine system undergoes oscillations, as seen in Fig.~\ref{fig:Conductance_QH_SC_SGM_spinless}(a). Let us now inspect the case of a slightly different value of the magnetic field, i.e., $B = 1$ T. Figure \ref{fig:G_and_SGM_pot_dist}(a) shows the conductance change map. As previously mentioned, we observe that as the tip moves closer to the SC interface, the conductance starts to oscillate rapidly. Importantly, as the bare conductance here (without the tip) $G_0 = 0.22\cdot 2e^2/h$ is far from the conductance quantum, the conductance change introduced by the tip can now vary from negative to positive. In the map, we also observe a broad region of amplified conductance without oscillations. This occurs even when the tip is far from the interface. We can understand this phenomenon by looking at the SGM tip potential distribution shown in Fig.~\ref{fig:G_and_SGM_pot_dist}(b). Here, we consider the tip localized at $x_\mathrm{tip} = - 800$ nm. Due to the non-zero tail of the potential close to the interface, the effective chemical potential at the position of the CAES state is slightly decreased (by roughly 0.05 meV), resulting in a positive shift of the conductance along the oscillatory pattern in $\mu$---see Fig. \ref{fig:G_and_SGM_pot_dist}(c).

\subsubsection{Explanation of the SGM induced conductance oscillations}
\begin{figure}
    \centering
    \includegraphics[width=0.85\linewidth]{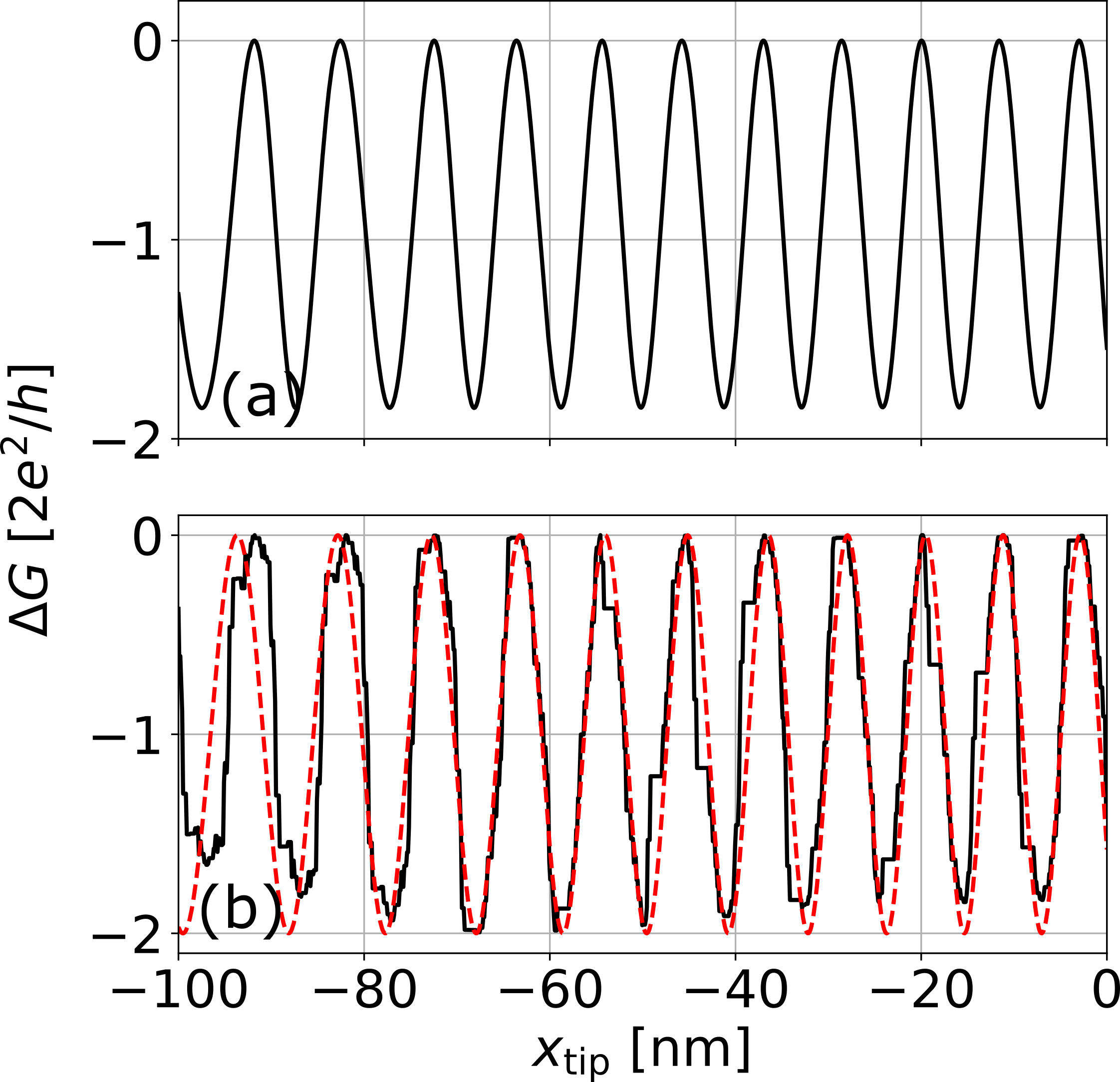}
    \caption{(a) Conductance cross-section from Fig. \ref{fig:Conductance_QH_SC_SGM_spinless}(b) for $y_\mathrm{tip} = 0$. (b) Black curve shows the conductance versus the tip position for the approximation of the circular hardwall potential. Red dashed curve shows the analytical approximation of the conductance due to Aharonov-Bohm interference obtained through Eq. \ref{analyticalG}. $\mu = 5$ meV and $B = 0.912$ T for all plots.}
    \label{fig:Conductance_crossection_SGM_loren_circular}
\end{figure}

The conductance maps Figs. \ref{fig:Conductance_QH_SC_SGM_spinless}(b) and \ref{fig:G_and_SGM_pot_dist}(a) show a unique feature of SGM mapping of the CAESs---as the tip potential starts reaching the right edge of the system, the conductance exhibits rapid oscillations. To inspect the conductance oscillations, we plot the cross-section of the map of Fig. \ref{fig:Conductance_QH_SC_SGM_spinless}(b) in Fig. \ref{fig:Conductance_crossection_SGM_loren_circular}(a). To simplify the analysis of the oscillations, let us approximate the smooth Lorentzian potential introduced by the tip at the 2DEG level with a hard-wall circular potential of amplitude 0.1~eV and radius $R_{\mathrm{hw}} = 110$ nm, with its center positioned at $(x_\mathrm{tip}, 0)$. The obtained conductance trace is shown in Fig. \ref{fig:Conductance_crossection_SGM_loren_circular}(b) with a black curve. We observe that it reproduces the trace obtained with the Lorentzian potential shown in Fig. \ref{fig:Conductance_crossection_SGM_loren_circular}(a).

\begin{figure*}
    \centering
    \includegraphics[width=0.98\linewidth]{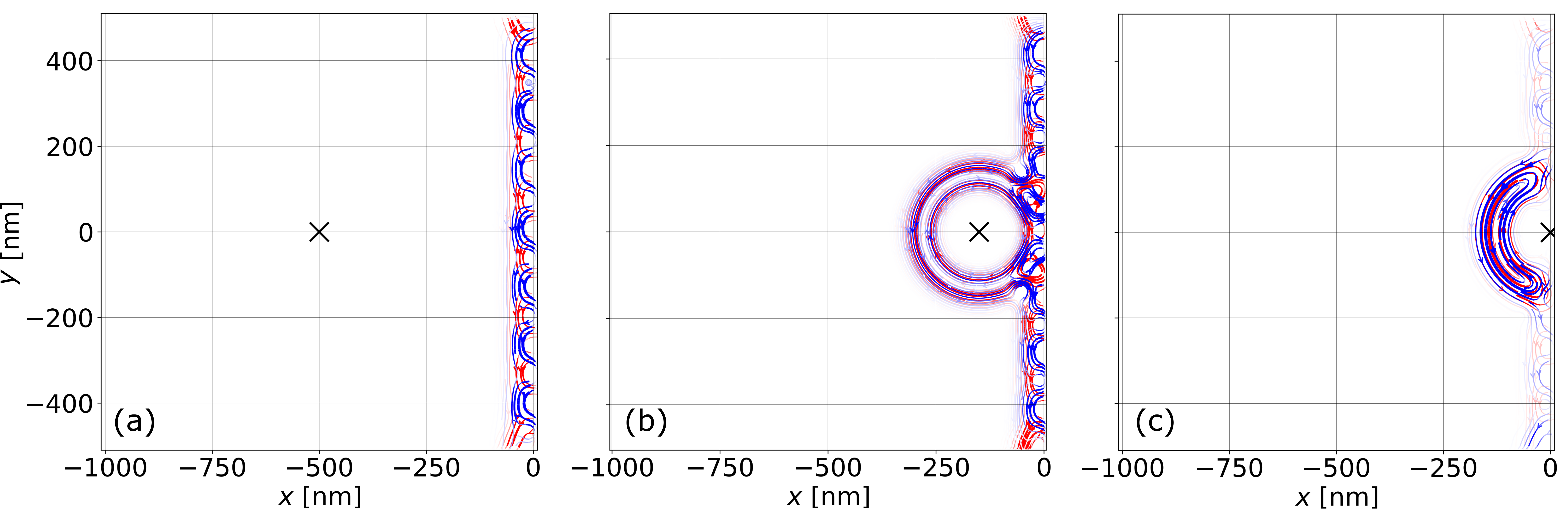}
    \caption{Probability currents obtained with SGM tip at magnetic field $B_z  = 0.912$ T and $\mu = 5$ meV (a) at $x_\mathrm{tip} = -500$ nm, (b) $x_\mathrm{tip} = -150$ nm, $y_{tip} = 0 $ (c) $x_\mathrm{tip} = 0$ nm, $y_{tip} = 0 $. The red (blue) colors correspond to electron (hole) part of the probability currents.}
    \label{fig:Current}
\end{figure*}

Figure \ref{fig:Current} shows the probability current carried by the CAES at the interface for the case of the approximation of a circular hard-wall SGM tip potential. The color of the current indicates the electron (red) or hole (blue) contribution. For the tip located at $x_{\mathrm{tip}} = -500$ nm, we observe that the SGM does not perturb the electron-hole flow along the NS interface. The probability current along the interface varies between the dominant electron and hole components, as expected for two CAESs. When the tip gets close to the NS interface, it decouples the modes from the edge, and due to the impossibility of backscattering, the electrons and holes start to flow around the potential introduced by the tip [Fig. \ref{fig:Current}(b)]. As the tip moves even further towards the interface, the electron and hole states are fully decoupled from the interface; the Andreev reflection at the decoupled trajectory stops, and consequently, we do not observe electron-hole oscillations on the tip determined trajectory [Fig. \ref{fig:Current}(c)]. 

When this tip is near the interface, it locally decouples the two CAESs into independent electron and hole edge states, which are forced to detour around the potential in the normal region. During the detour in the normal region, the electron and hole states do not mix, but they accumulate a relative phase due to the  magnetic field that, as we show in the following, results in Aharonov-Bohm-type oscillations.

The two CAESs at the NS interface ($\Psi_+$ and $\Psi_-$) are superpositions of the pure electron $|e\rangle$ and hole $|h\rangle$ states. We find the electron and hole contributions to be approximately equal and define the normalized CAESs basis as:
\begin{equation}
    \Psi_+ = \frac{1}{\sqrt{2}} \begin{pmatrix} 1 \\ 1 \end{pmatrix} = \frac{1}{\sqrt{2}} (|e\rangle + |h\rangle),
\end{equation}
\begin{equation}
    \Psi_- = \frac{1}{\sqrt{2}} \begin{pmatrix} 1 \\ -1 \end{pmatrix} = \frac{1}{\sqrt{2}} (|e\rangle - |h\rangle).
\end{equation}

When a circular scattering potential centered at $(x_\mathrm{tip}, 0)$ approaches the interface, it pushes the wave functions radially outward into the normal region with $R_\mathrm{eff}$, and the effective pairing potential smoothly drops to zero ($\Delta \to 0$). This evolution of the pairing potential results in an adiabatic change of CAESs into the pure normal-state basis states $\Psi_+ \to |e\rangle, \Psi_- \to |h\rangle$. 

The decoupled electron and hole travel in the normal region ($x < 0$) around the circular potential, whose boundary is given by $(x - x_\mathrm{tip})^2 + y^2 = R_\mathrm{eff}^2$. The total enclosed area is
\begin{equation}
    \Omega(x_\mathrm{tip}) = R_\mathrm{eff}^2 \arccos\left(\frac{x_\mathrm{tip}}{R_\mathrm{eff}}\right) - x_\mathrm{tip}\sqrt{R_\mathrm{eff}^2 - x_\mathrm{tip}^2}.
    \label{eq:area}
\end{equation}

The magnetic phase accumulated by a charge $q$ over path $C$ is $\phi = \frac{q}{\hbar} \int_C \vb{A} \cdot d\vb{l}$, and for the assumed Landau gauge the integral becomes $-B \cdot \Omega(x_\mathrm{tip})$. The electron ($q = -e$) and hole ($q = +e$) accumulate respective phases, and the relative Aharonov-Bohm phase difference picked up when detouring the tip is
\begin{equation}
    \Delta\phi_\mathrm{AB} = 2\pi \frac{B \cdot \Omega(x_\mathrm{tip})}{\Phi_0},
\end{equation}
where $\Phi_0 = h/2e$ is the superconducting flux quantum.

Over the total remaining length of the NS interface (before and after the tip), the adiabatic evolution allows the $\Psi_{\pm}$ eigenmodes to propagate independently. They accumulate total phases $\Phi_+$ and $\Phi_-$, where the total phase difference is $\Delta\Phi = \Phi_+ - \Phi_- = \Phi_\mathrm{dyn} + \Delta\phi_\mathrm{AB}$. The dynamic phase $\Phi_\mathrm{dyn} = \int (k_+ - k_-) dy$ remains constant when the tip moves strictly in the $x$-direction and also in the $y$-direction, as the length before and after the tip that contributes to the geometrical phase remains constant (unless the tip potential is close to one of the normal leads).

To find the conductance, we trace an electron injected from the normal lead. Inverting our definitions of the CAESs basis, the incident electron state is $|e\rangle = \frac{1}{\sqrt{2}} \Psi_+ + \frac{1}{\sqrt{2}} \Psi_-$. This state propagates coherently to the bottom lead (drain), accumulating the channel-specific phases
\begin{equation}
    |\psi_\mathrm{drain}\rangle = \frac{1}{\sqrt{2}} e^{i\Phi_+} \Psi_+ + \frac{1}{\sqrt{2}} e^{i\Phi_-} \Psi_-.
\end{equation}
We project this final state onto the electron mode to find the exit amplitude, $\psi_{e, \mathrm{out}} = \langle e | \psi_\mathrm{drain} \rangle = \frac{1}{2} e^{i\Phi_+} + \frac{1}{2} e^{i\Phi_-}$. The probability $P_e$ of measuring an electron at the drain is the absolute square $|\psi_{e, out}|^2$, $P_e = \frac{1}{2} + \frac{1}{2} \cos(\Delta\Phi)$.

The normalized differential conductance $G$ relates to the net charge transfer probability ($P_e - P_h$). Due to the lack of backscattering $P_e + P_h = 1$, we have $G \propto 2P_e - 1$ and
\begin{equation}
    G(x_\mathrm{tip}) = 2 \left[ \frac{1}{2} + \frac{1}{2} \cos(\Delta\Phi) \right] - 1 = \cos(\Delta\Phi).
\end{equation}

Absorbing the constant dynamical phase into a baseline offset $\phi_0$, the exact analytical equation governing the conductance oscillations maps directly to the tip's spatial position:
\begin{equation}
    G(x_\mathrm{tip}) = \cos\left( \frac{2\pi B}{\Phi_0} \Omega(x_\mathrm{tip}) + \phi_0 \right).
\label{analyticalG}
\end{equation}

In Fig. \ref{fig:Conductance_crossection_SGM_loren_circular}(b) with the red curve, we plot the resulting analytical conductance obtained using formula Eq. \ref{analyticalG}, with fitted $R_\mathrm{eff}=137$ nm (which accounts for the finite thickness of the electron and hole stream, larger than $R_\mathrm{hw}$) and $\phi_0$. We observe a nearly perfect agreement with the numerical conductance, which proves that the conductance maps directly to the enclosed Aharonov-Bohm phase. Note that deviation from equal electron and hole contributions in CAESs would result in a decrease in the oscillation amplitudes; nevertheless, in our case, the oscillations have an amplitude close to $2 \cdot 2e^2/h$, which is consistent with the numerically obtained equal electron and hole contributions to CAESs.

\subsection{Higher filling factors}
\begin{figure}
    \centering
    \includegraphics[width=0.85\linewidth]{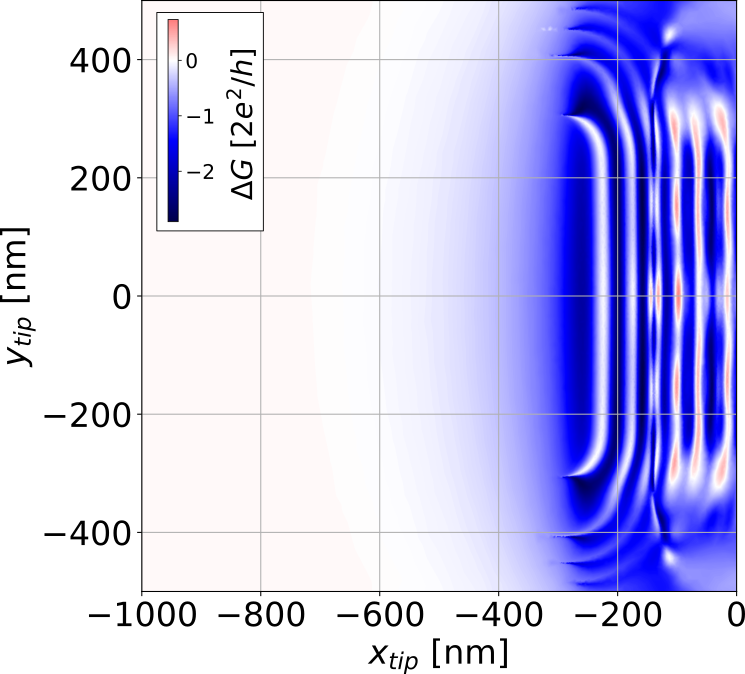}
    \caption{SGM conductance change plot versus the tip position for $B = 0.343$ T and $\mu = 5$ meV with $\nu = 4$.}
    \label{fig:Conductance_nu_4}
\end{figure}

\begin{figure}
    \centering
    \includegraphics[width=0.85\linewidth]{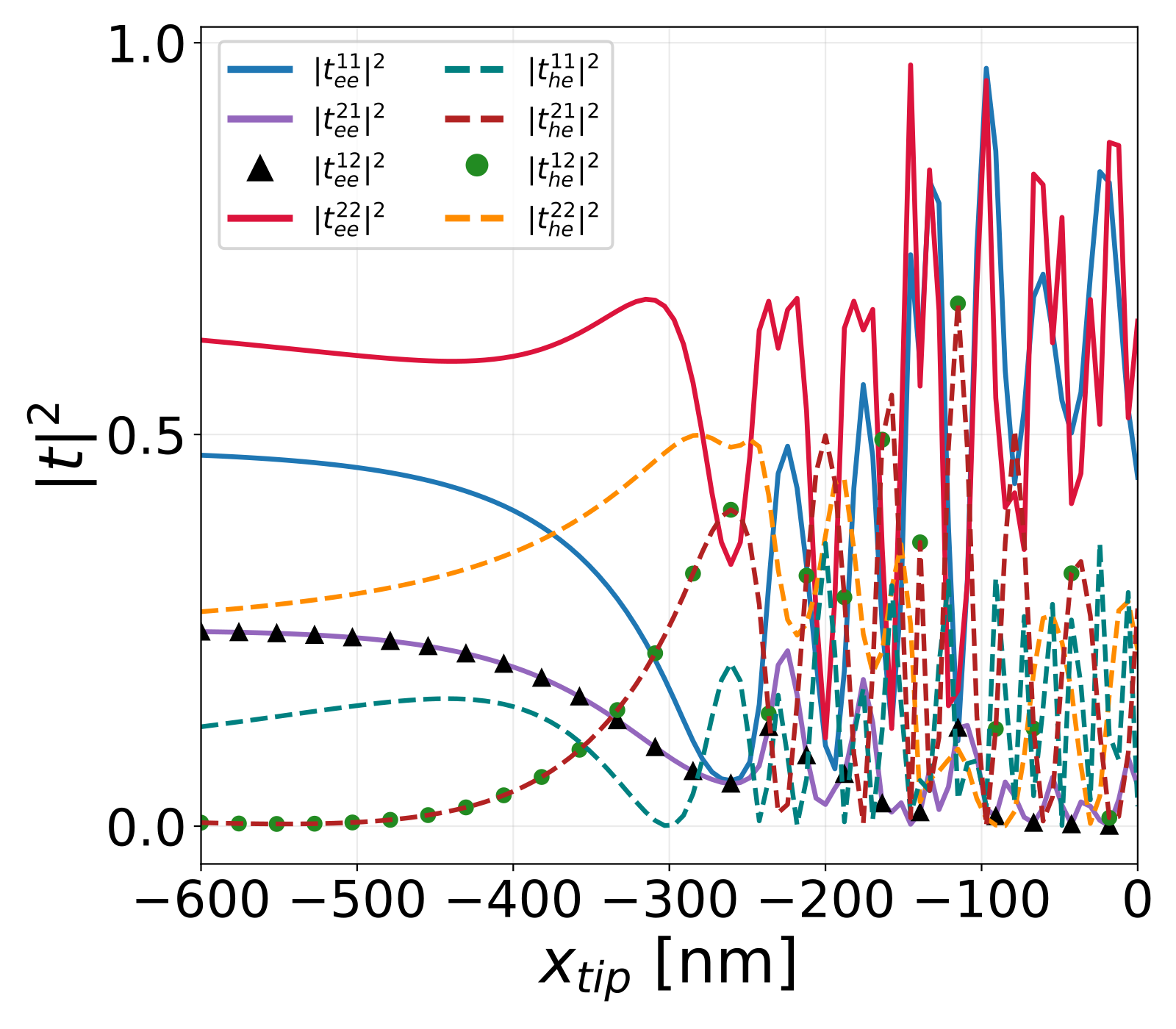}
    \caption{Components of electron and hole transmission probabilities in the $\nu = 4$ regime. The perpendicular magnetic field is $B_z = 0.343$ T and $\mu = 5$ meV.}
    \label{fig:trnasmission_probability}
\end{figure}

Let us now inspect the $\nu = 4$ case. The corresponding conductance change map is shown in Fig. \ref{fig:Conductance_nu_4}. Similar to the lowest filling factor case, we observe conductance oscillations as the tip approaches the SC interface. Here we deal with two pairs of CAESs corresponding to the two  electron (hole) edge modes. Without the superconductor, the electron QH edge modes propagate independently. In the presence of the SC interface, Andreev reflection leads to their mixing, as we showed in our previous paper \cite{maji2026}. This mixing can be inspected in the electron to electron and electron to hole transmission probabilities displayed in Fig. \ref{fig:trnasmission_probability}. For more than a single edge channel in the normal lead, they are defined as $|t_{ee}^{ij}|^2$, $|t_{eh}^{ij}|^2$, where $i,j \in \{1,2\}$ correspond to the output and input quantum edge mode channels, respectively.

When the tip is far from the interface, there is a clear signature of mode mixing; e.g., the electron entering the system in the first edge mode escapes the system intermixed between the two edge modes (see the violet curve). Nevertheless, the contributions of intermode transport are smaller than those corresponding to intramode transport. When the SGM tip approaches the interface, it not only introduces Aharonov-Bohm interference but also strongly affects the mode mixing, intensifying the redistribution of the transmission probability between the modes.

\begin{figure*}
    \centering
    \includegraphics[width=0.85\linewidth]{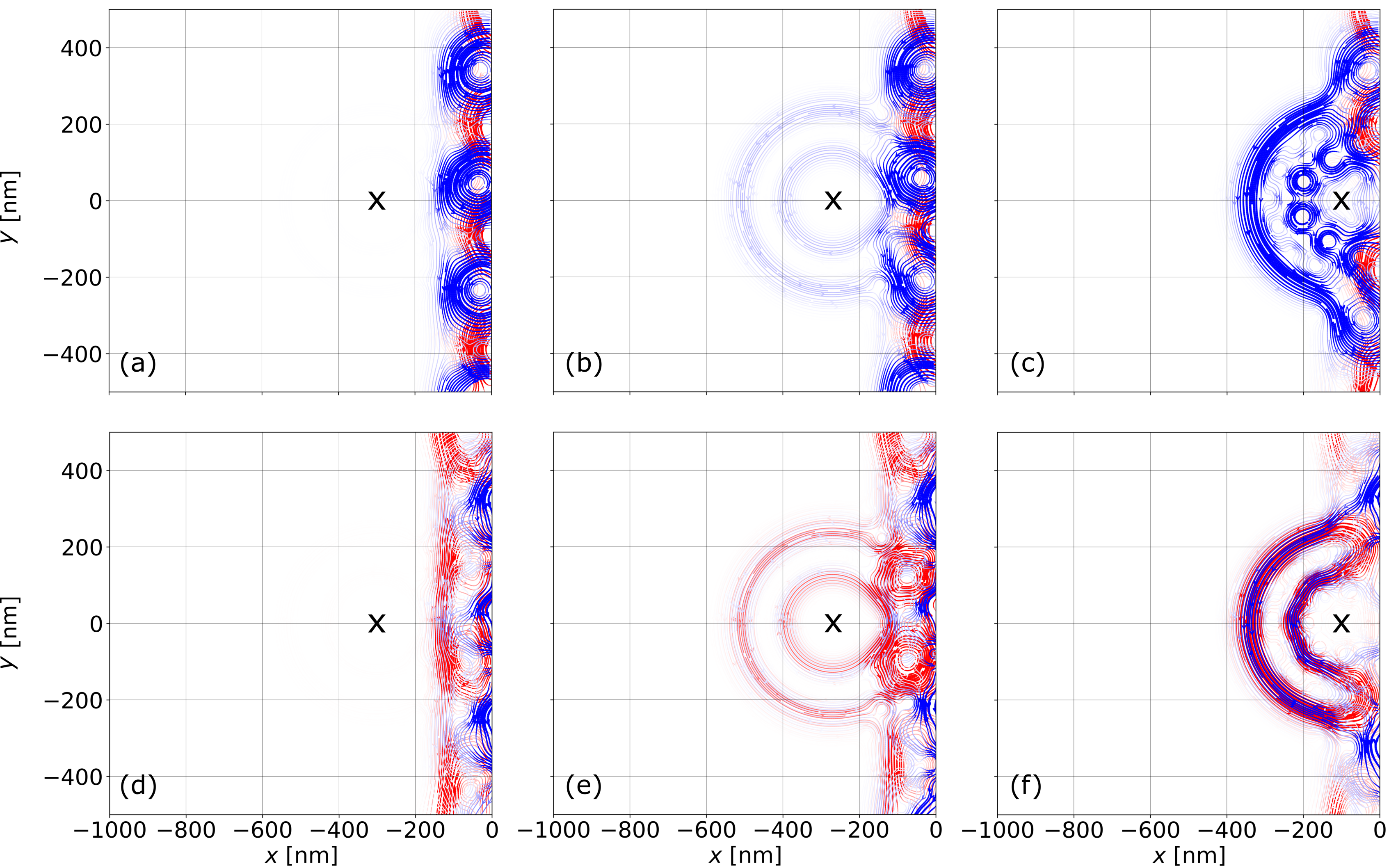}
    \caption{Probability currents obtained for $\nu = 4$, $B = 0.343$ T and $\mu = 5$ meV with SGM tip at (a), (d) $x_\mathrm{tip} = -300$ nm, $y_\mathrm{tip} = 0 $; (b), (e) $x_\mathrm{tip} = -270$ nm, $y_\mathrm{tip} = 0 $; (c), (f) $x_\mathrm{tip} = -100$ nm, $y_\mathrm{tip} = 0$. The upper panel contains the probability current for CAESs for the electron injected in the first edge mode; the lower panel contains the current of the electron injected in the second edge mode.}
    \label{fig:total_Current}
\end{figure*}

In the map of Fig. \ref{fig:Conductance_nu_4}, we observe two regimes with different patterns of oscillations. The two electron edge modes have different spatial spans along the $x$ direction, and as the tip moves towards the SC interface, it first starts affecting the one with the larger span in the $x$ direction, inducing an Aharonov-Bohm loop in it. This can be observed in the current maps of Fig. \ref{fig:total_Current}, where in the top row we present the current resulting from the electron being injected in the first edge mode (with a smaller spatial span), and in the second row, the current resulting from the electron being injected in the second edge mode (with a larger spatial span). We observe that as the tip approaches the interface, for $x_\mathrm{tip} = -270$ nm, the current resulting from the injection in the first edge mode remains largely unaffected---leaving pronounced electron-hole oscillations at the interface [cf. Figs. \ref{fig:total_Current}(a) and (b)],  while the current resulting from the electron injected in the second mode exhibits strong perturbation by the tip [cf. Figs. \ref{fig:total_Current}(d) and (e)]. When the tip moves closer to the edge, it strongly affects the current running through both edge modes, effectively decoupling them from the interface [see Figs. \ref{fig:total_Current}(c) and (f)] and amplifying the Aharonov-Bohm interference observed in Fig. \ref{fig:trnasmission_probability} for $x_\mathrm{tip} < 170$ nm.

\subsection{SGM mapping of a disordered system}
\begin{figure}
    \centering
    \includegraphics[width=0.85\linewidth]{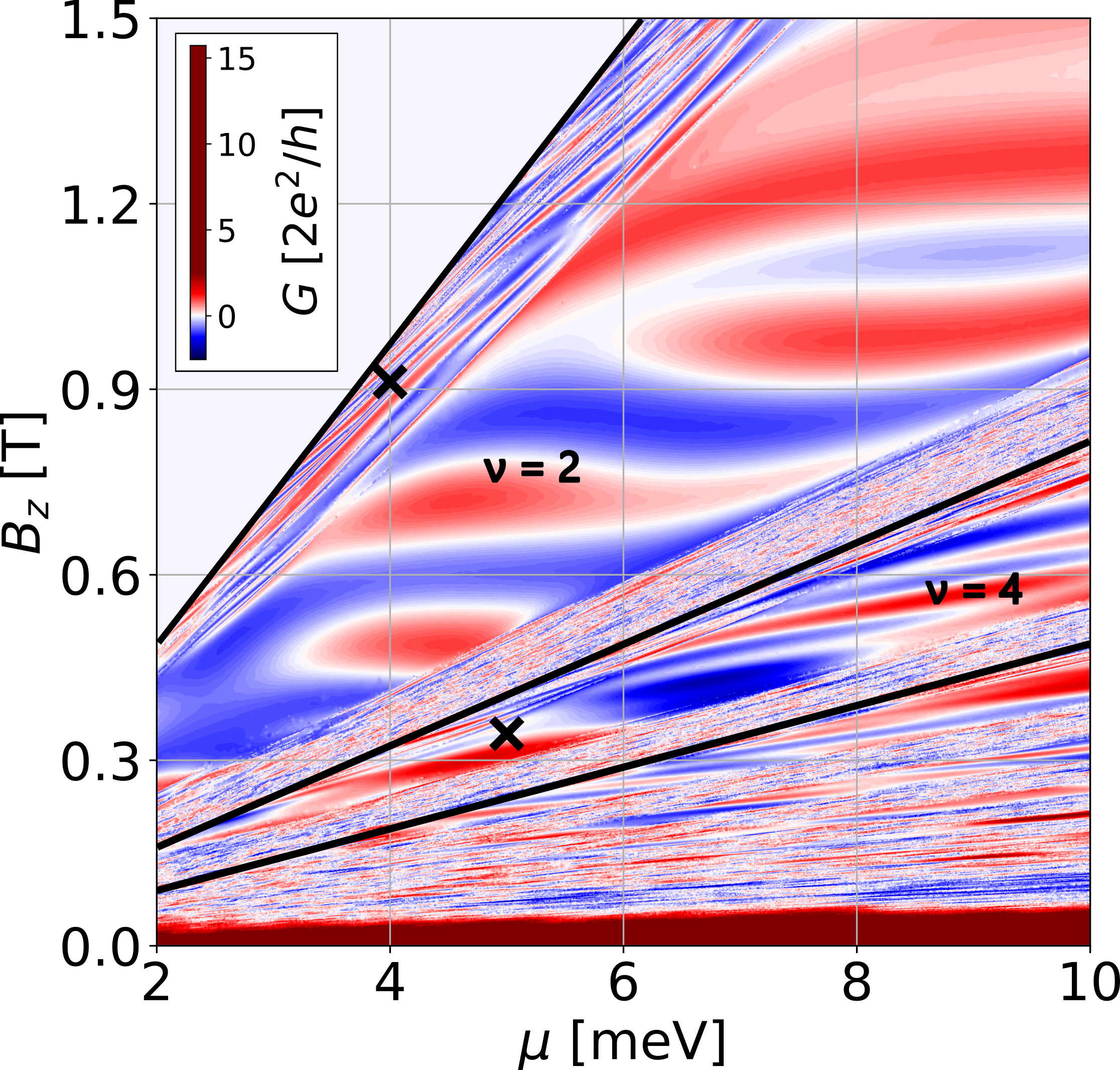}
    \caption{Conductance map versus chemical potential and the magnetic field with disorder with $l_e= 500$ nm. The crosses denote parameter points for which the SGM conductance change maps are obtained in Fig. \ref{fig:Conductance_SGM_1000_2000}(c) and (d).}
    \label{fig:Disorder_conductance_le}
\end{figure}

In realistic QH-SC devices, disorder is common and may arise from device impurities, interfacial imperfections, or electrostatic inhomogeneities that can affect the formation or propagation of CAESs. In a 2DEG-SC system, disorder can modify the magnetotransport and the quantum interference effects \cite{Takagaki2020Disorder}. Disorder near the QH-SC interface can also affect the Andreev conversion and the corresponding transport response \cite{Kurilovich2023Disorder, Antonio}. To assess the possibility of using SGM as a tool to detect CAESs in disordered systems, we consider the QH 2DEG system with a limited mean free path. The finite mean free path is implemented through the inclusion of an onsite potential uniformly distributed within the range $[-U_d/2, U_d/2]$, with the amplitude derived for magnetotransport calculations in a discretized model of a two-dimensional system\cite{Ando}
\begin{equation}
    U_d=\mu \sqrt{\frac{6\lambda_F^3}{\pi^3a^2l_e}}.
\end{equation}
Here, $a$, $\lambda_F$, and $l_e$ are the lattice constant, the Fermi wavelength, and the mean free path, respectively. We treat the disorder amplitude as a constant obtained with $\lambda_F=2\pi\hbar/\sqrt{2m^*\mu}$, for $\mu = 5$ meV.

\begin{figure*}
    \centering
    \includegraphics[width=0.98\linewidth]{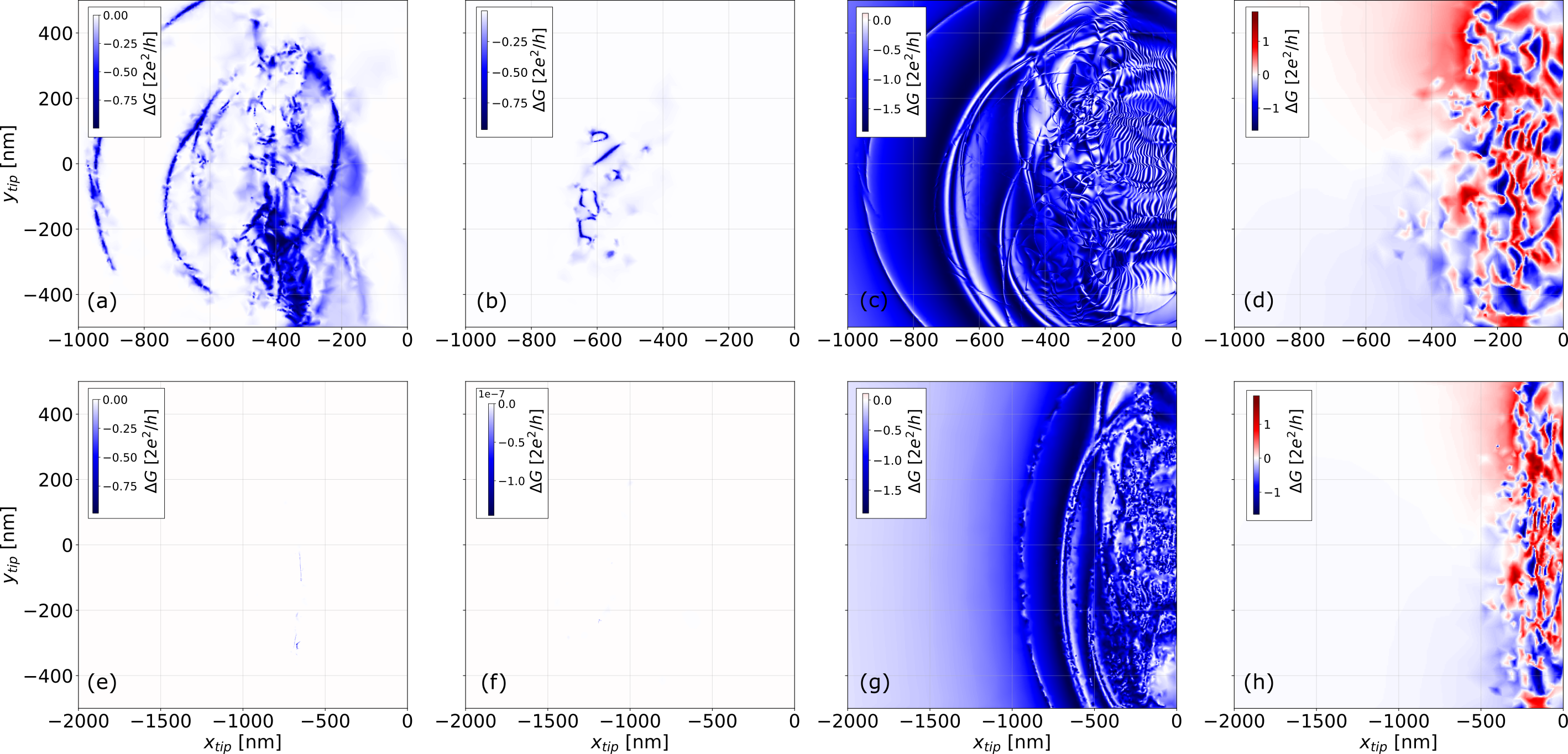}
    \caption{Conductance change plot for disordered system with $l_e = 500$ nm with respect to the SGM tip position. The first row corresponds to the system with $W = 1000$ nm and the second row to a system with $W = 2000$ nm. First and second columns of the plot (a), (b), (e), (f) correspond to the normal system. The third and fourth columns, (c), (d), (g), (h) correspond to the normal system with a SC interface. (a), (e), (c), (g) are plotted at $B = 0.912$ T for $\nu = 2$ at $\mu = 4$ meV and Fig. (b), (f), (d), (h) are obtained at $B = 0.343$ T for $\nu = 4$ at $\mu = 5$ meV.}
    \label{fig:Conductance_SGM_1000_2000}
\end{figure*}

In Fig. \ref{fig:Disorder_conductance_le}, we show the conductance map obtained for a disordered system versus the chemical potential and the magnetic field. We observe a strong disturbance of the soft oscillating patterns displayed previously in Fig. \ref{fig:Conductance_QH_SC_SGM_spinless}(a) for a clean system. The strongest effect of the disorder is observed in the vicinity of each $\nu$ sector line, where the disorder-broadened Landau levels cross the Fermi energy, leading to a strong admixture of CAESs. Despite the lack of clear CAESs oscillations in the conductance map for given filling factors as in the case of a clean system, we show that, particularly in this dirty case, the SGM technique is able to validate the presence of CAESs.

First, let us look at the maps of conductance change obtained for a purely normal system. We plot the conductance change in Fig. \ref{fig:Conductance_SGM_1000_2000}(a) and (b) for the cases of $\nu = 2$ and $\nu = 4$, respectively, and for our system with the width of $W = 1000$ nm. Since there is considerably strong disorder, the SGM tip introduced into the system can induce backscattering, as the tip and the  disorder introduce a bridge that connects the two edges of the system and allows electrons entering the system to be backscattered on the other edge---see Fig. \ref{fig:Conductance_SGM_1000_2000}(a-b). The SGM conductance change diminishes when this bridge is weakened, which is the case for a system with a larger distance between the edges, presented in Fig. \ref{fig:Conductance_SGM_1000_2000}(e) and (f) for elevated $W = 2000$ nm. There we observe that the conductance is independent of the position of the tip, as the backscattering is muted. 

Importantly, for the two widths, the conductance change map in the presence of CAESs reveals significant features at the tip position close to the edge at which CAESs propagate. This is true for both $\nu = 2$ [shown in Figs. \ref{fig:Conductance_SGM_1000_2000}(c) and (g)] and $\nu = 4$ [shown in Figs. \ref{fig:Conductance_SGM_1000_2000}(d) and (h)]. Obviously, for this disordered case, the clean Aharonov-Bohm periodic oscillations that were previously constant along the $y$-direction are lost, as the charge carrier orbits are significantly randomized. However, the conductance response of the system clearly reveals the presence of an edge state and the interference of the electron and hole parts of the wave function. Furthermore, the distinction between the $\nu = 2$ and $\nu = 4$ cases is possible by monitoring the amplitude of the oscillating conductance that corresponds to multiples of the conductance quantization constant; for the $\nu = 2$ case, it varies within $|\Delta G| < 2\cdot2e^2/h$, and for $\nu = 4$, it exceeds this value and with $|\Delta G| < 4\cdot2e^2/h$. 

\section{Summary and Conclusions}
We theoretically investigated the spatial responses of chiral Andreev edge states at a quantum Hall-superconductor interface, probed by scanning gate microscopy using the tight-binding Bogoliubov-de Gennes formalism. In the lowest filling factor regime ($\nu=2$), when the scanning gate tip approaches the superconductor, it locally decouples the chiral Andreev edge states into independent electron and hole edge states. Detoured around the tip-induced potential in the normal region, these decoupled quasiparticles accumulate a relative magnetic phase before recombining at the superconductor, yielding rapid Aharonov-Bohm-type conductance oscillations—a spatial response strictly absent when probing unidirectionally propagating standard quantum Hall edge modes. We demonstrated that these oscillations are  captured by a geometric analytical model based on the modified magnetic flux enclosed by the detoured trajectories. Extending our analysis to higher filling factors (e.g., $\nu=4$), the localized tip potential fundamentally alters the preexisting Andreev reflection-induced mode mixing. Because the spatially distinct edge modes possess different spatial spans, the tip interacts with them progressively, first inducing localized Aharonov-Bohm loops in the outermost mode before decoupling both channels from the interface entirely. This progressive spatial perturbation actively redistributes transmission probabilities among the edge channels, driving complex hybridization and amplifying the transport oscillations beyond simple two-mode interference. Finally, we observed robust spatial signatures of chiral Andreev edge states even in the presence of disorder. We showed that while a finite mean free path destroys the perfectly periodic conductance oscillations seen in pristine systems, the scanning gate microscopy technique still successfully extracts a strong, highly localized interferometric response near the interface. These unique local conductance variations serve as direct signatures capable of revealing both the presence of hybrid electron-hole states and the system's underlying filling factor in non-ideal experimental environments.

\section*{Acknowledgments}
This work was supported by the National Science Center, Poland (NCN) agreement number UMO-2020/38/E/ST3/00418. We gratefully acknowledge the Polish high-performance computing infrastructure PLGrid (HPC Center: ACK Cyfronet AGH) for providing computer facilities and support within the computational grant no. PLG/2026/019695.
\bibliography{references}
\end{document}